# The Emergence of Topological Nodal Points in Photonic Crystal with Mirror Symmetry

Wen-Yu He and C. T. Chan*
*Department of Physics and Institute for Advanced Study,*
*Hong Kong University of Science and Technology,*
*Clear Water Bay, Kowloon, Hong Kong, China*

We show that topological nodal points can emerge in photonic crystal possessing mirror symmetry. The mechanism of generating topological nodal points is discussed in a two-dimensional photonic square lattice, in which four topological nodal points split out naturally after the touching of two bands with different parity. The emergence of such nodal points, characterized by vortex structure in momentum space, is attributed to the unavoidable band crossing protected by mirror symmetry. The topological nodes can be unbuckled through breaking the mirror symmetry and a photonic Chern insulator can be achieved through time reversal symmetry breaking. The joint effect of breaking time reversal symmetry and breaking inversion symmetry is further found to strengthen the finite size effect, providing ways to engineer helical edge states.

*Introduction.*—Topologically characterized gapless points induce many novel phenomena in both electronic and photonic systems, from robust edge states to back-scattering immune transport [1-9]. At the band touching point, vortex structure for Dirac nodal points or monopole for Weyl nodes appear in the momentum space as the topological feature. As the remarkable properties are mainly attributed to the topological nature of these nodal points, the mechanism of the creation, moving and vanishing of them, attracts significant research attention both theoretically and experimentally [6-7, 10-13]. In previous studies, the creation and manipulation of topological nodal points is inevitably involved with symmetry operation. For Dirac nodes in honeycomb or square lattice, lattice anisotropy is necessary to induce the topological transition [6, 11-13], while for Weyl nodes in double gyroid photonic crystals, either parity or time reversal symmetry breaking is required [7]. However, interesting band degeneracy phenomenon also occurs through variation of parameters unrelated to symmetry, as the exotic topological semimetal phase in Ref. 8 and the photonic Dirac cone at Brillouin zone center in Ref. 15. Such results motivate us to explore the possibility to produce and manipulate topological nodal points in an unconventional way.

In this paper, we show theoretically that topological nodal points can potentially emerge in all photonic crystal with mirror symmetry during the closing of a gap. When the mirror symmetry is present, the eigenmodes along the high symmetry line that are invariant under the mirror operator can be classified by the mirror symmetry representation. Once two bands with different parity are tuned to approach each other, band crossing is unavoidable after their touching. Such unavoidable crossing induced band degeneracy is stable and cannot be destroyed unless the mirror symmetry is broken. We investigate in detail the property of such band degeneracy in a photonic square lattice comprising dielectric cylinders. The photonic square lattice possesses two equivalent mirror planes, generating two pairs of topological nodal points as expected. With k·p perturbation and symmetry analysis using group theory, we obtain the effective Hamiltonian near degeneracy and present the evolution of subbands to show the mechanism of generating nodal points. Subsequently the effect of mirror symmetry breaking is studied through replacing dielectric cylinders with inversion asymmetric artificial structure. We found that fixing its principal axis parallel to one of the mirror planes would lift one pair of nodal points and preserve the other, while a general orientation would break the two, as is consistent with its mirror symmetry protection. The topological feature of the nodal points is manifested by considering magneto-optical effect. A photonic Chern insulator with two helical edge states is realized due to the breaking of time reversal symmetry. Furthermore, introducing inversion asymmetric artificial structure to such photonic Chern insulator, the finite size effect [16] is found to be strengthened, which suppresses one helical edge state while does not influence the other at specific frequency.

*Topological nodal points and mirror symmetry.* — Consider a photonic crystal possessing mirror symmetry, such mirror reflection invariance gives

$$MH(\mathbf{k})M^{-1} = H(R\mathbf{k})$$

where $M$ is the mirror operator and $R$ is the 2×2 mirror reflection matrix defining the mirror reflection in a 2D plane. In the Brillouin zone, mirror reflection invariant line satisfies the condition $R\mathbf{k}=\mathbf{k}+\mathbf{G}$ (where $\mathbf{G}$ is reciprocal lattice vector). Along the mirror reflection invariant line, the eigenfunction of the Hamiltonian is also an irreducible representation of the mirror operator, and thus all bands along that direction can be labeled by {A, B}, which correspond to the representation with even and odd parity respectively under the mirror operator. If two bands with different parity get close to each other, the effective Hamiltonian can be approximated on the basis of A and B.

On this basis, the matrix representation of $M$ is $\sigma_z$. Along the mirror reflection invariant line, the symmetry constraint upon the effective Hamiltonian makes its off diagonal elements become zero and only leaves diagonal terms. Such diagonal terms control the dispersion of the two bands independently, and the two bands can have linear crossing unavoidably once after their touching, as is shown in Fig. 1. Near the crossing point, expanding the effective Hamiltonian to the first order in the rule of symmetry constraint, we can generate the effective Hamiltonian as the following

$$H_{eff}(k_\parallel, k_\perp) = \frac{1}{2}(v_A + v_B)k_\parallel \sigma_0 + v_+ k_\perp \sigma_+ + v_- k_\perp \sigma_- + \frac{1}{2}(v_A - v_B)k_\parallel \sigma_z$$

where $k_\parallel$ and $k_\perp$ represent the vector component parallel and perpendicular to the mirror reflection invariant line respectively, and $\sigma_\pm = \sigma_x \pm i\sigma_y$; the detailed illustration about the symmetry constraint is relegated in the Supplemental Material [17]. Therefore the crossing point has a linear dispersion in all direction and acts as a source (sink) of Berry curvature in the momentum space. When the dielectric constant of cylinders is very high, a photonic crystal has band gaps separating isolated bands. As we progressively reduce the dielectric constant, the photonic bands will close as some bands separated by the band gaps will approach and touch each other. In this process, if mirror symmetry is present, touching between photonic bands in the mirror reflection invariant line with different parity will make topological nodal points split out naturally. This intrinsic connection in principal provides the possibility for the emergence of topological nodal points in photonic crystal with mirror symmetry.

*Emergence of topological nodal points.* —We consider a two-dimensional photonic crystal system in a square lattice consisting of dielectric cylinders. We perform plane wave expansion [18] to calculate the band structure for the transverse magnetic (TM) polarization with the electric field along the rod axis. Here we denote $a$ as the lattice constant, and relative permittivity and radius of the cylinders are $\varepsilon=5.4$, $r=0.2a$ respectively. Along the XM direction, the fourth and fifth bands cross linearly at frequency $\omega_0$, as is shown in Fig. 2(a), suggesting that isolated band degeneracy occurs in the momentum space. Further calculation of three-dimensional band dispersion demonstrates that band touching with linear dispersion arises around X and Y as shown in Fig. 2(c) and (d), giving rise to four gapless points in the first Brillouin zone. Emergence of nodal points can also appear in other lattices as long as there is a mirror symmetry. For example, six topological nodal points can also emerge in a photonic crystal with triangular lattice as can be seen in Fig. S1 in [17]. This phenomenon is stable and immune to small variation in radius and permittivity of rods, which can only shift degenerate points in frequency and momentum space but cannot remove them. The band degeneracy vanishes once a pair of degeneracy points encounter at X (Y), where they split out, and pairwise annihilation occurs. Such characteristics of the degeneracy points resemble the behaviors of Weyl fermions [19, 20], indicating the topological nature of the nodal points.

*$k \cdot p$ perturbation and symmetry protection.* —The Maxwell equation for TM wave in two dimensional photonic crystals can be converted to an eigen problem. We can express the Bloch wave function $\psi_{n\mathbf{k}}$ as a linear combination of $\psi_{n\mathbf{k_0}}$ with eigenfrequency $\omega_{n0}$, and establish an effective Hamiltonian with matrix element $\langle \psi_i | \hat{H} | \psi_j \rangle$ to describe the photonic system [21]. For the square lattice, all symmetry operations in the $C_{2V}$ group will keep the X (Y) point invariant, suggesting that the symmetry type of eigenfunction at X (Y) belongs to {$B_2$, $A_1$, $B_1$, $A_2$}, the irreducible representations of $C_{2V}$ [22]. For this reason we adopt $\Psi = (\psi_{B_2}, \psi_{A_1}, \psi_{B_1}, \psi_{A_2})$, which corresponds to the eigenfunction for the third, fourth, fifth, and sixth band respectively, as basis wave function [23] to do $k \cdot p$ perturbation near X point. Then a $4 \times 4$ effective Hamiltonian can be deduced as following

$$H = \begin{pmatrix} \frac{\omega_{B_2}^2}{c^2} + q_{B_2 B_2}\kappa^2 & -p_{B_2 A_1 y}\kappa_y & 0 & -p_{B_2 A_2 x}\kappa_x \\ -p_{B_2 A_1 y}^*\kappa_y & \frac{\omega_{A_1}^2}{c^2} + q_{A_1 A_1}\kappa^2 & -p_{A_1 B_2 x}\kappa_x & 0 \\ 0 & -p_{A_1 B_2 x}^*\kappa_x & \frac{\omega_{B_1}^2}{c^2} + q_{B_1 B_1}\kappa^2 & -p_{B_1 A_2 y}\kappa_y \\ -p_{B_2 A_2 x}^*\kappa_x & 0 & -p_{B_1 A_2 y}^*\kappa_y & \frac{\omega_{A_2}^2}{c^2} + q_{A_2 A_2}\kappa^2 \end{pmatrix}$$

Here $\boldsymbol{\kappa} = \mathbf{k} - \mathbf{k_0}$ is set, c is the speed of light, $\omega_i$ ($i \in \{B_2, A_1, B_1, A_2\}$) is the eigen frequency at X for corresponding band and the coefficient is

$$\mathbf{p}_{lj} = i\frac{(2\pi)^2}{\Omega} \int \psi_{l\mathbf{k_0}}^*(\mathbf{r}) \cdot \left\{ \frac{2\nabla \psi_{j\mathbf{k_0}}(\mathbf{r})}{\mu(\mathbf{r})} + \left[\nabla \frac{1}{\mu(\mathbf{r})}\right] \psi_{j\mathbf{k_0}}(\mathbf{r}) \right\} d\mathbf{r}$$

and

$$q_{lj} = \frac{(2\pi)^2}{\Omega} \int \psi_{l\mathbf{k_0}}^*(\mathbf{r}) \frac{1}{\mu(\mathbf{r})} \psi_{j\mathbf{k_0}} d\mathbf{r},$$

which can be obtained by numerical integration of the eigen wave function through the unit cell. The $\Omega$ here is the area of the unit cell to normalize the wave function. Results from this effective Hamiltonian agree well with plane wave calculation as can be seen from Fig. 2 (b). From group theory, this effective Hamiltonian should keep invariant under any symmetry operation of $C_{2V}$, which means that nonzero Hamiltonian matrix element only exists when the direct product of the irreducible representations of $\psi_i$, $\hat{H}$, and $\psi_j$ contains $A_1$, the full symmetry representation [24]. It can be confirmed that this effective Hamiltonian is consistent with the requirement by group theory [17], indicating the emergence of such topological nodal points is intrinsically decided by the crystal symmetry.

From the effective Hamiltonian, we see that it is block diagonalized along ΓX and XM direction. For $\kappa_x=0$ along XM, $A_1$ state only couples with $B_2$ state while $B_1$ state only couples with $A_2$ state, reducing $A_1$ state and $B_1$ state to A and B type respectively (here A and B are irreducible representation of mirror operator $m_x$). As $A_1$ and $B_1$ bands approach with variation of permittivity in cylinders, band crossing would unavoidably occur along XM. Such unavoidably crossing is intrinsically protected by the mirror symmetry of $m_x$, and guarantees the emergence of degeneracy at $N_1$ and $N_2$. However, for $\kappa_y=0$ along ΓX, $A_1$ and $B_1$ state are both reduced to A type. Thus in the process of band evolution, mutual repelling and inversion [17] between the two A type bands takes place once they touch. Such band inversion and repelling induces band gap along ΓX, acting as the role of partial gap to isolate the two band crossing points and making a pair of nodal points $N_1$ and $N_2$ split out, as is shown in Fig. 3. Here we see, the nodal points $N_1$ and $N_2$ are stable and generic once they emerge, as long as the $m_x$ mirror symmetry protected unavoidable crossing maintains. Such analysis is also suitable for nodal points $N_3$ and $N_4$ near Y, which are protected by $m_y$ mirror symmetry.

Since the nodal points are formed by linearly crossing of the fourth and fifth bands, far away from the third and sixth bands in frequency, we further use second order perturbation to integrate out the distant frequency degrees of freedom and reduce the effective Hamiltonian to a two band model as [17]

$$\tilde{H} = \frac{1}{2}\left(a(\kappa_x,\kappa_y)+c(\kappa_x,\kappa_y)\right)I + b(\kappa_x,\kappa_y)\sigma_y + \frac{1}{2}\left[a(\kappa_x,\kappa_y)-c(\kappa_x,\kappa_y)\right]\sigma_z$$

where $\sigma_y$ and $\sigma_z$ are two Pauli matrices and $I$ is the 2×2 identity matrix. Using the coefficients of Pauli matrices to define a planar vector $\mathbf{h}=(h_x,h_y)$, with $h_x=b(\kappa_x, \kappa_y)$ and $h_y=1/2[a(\kappa_x, \kappa_y)-c(\kappa_x, \kappa_y)]$, we can interpret the nodal points as a topological vortex, with topological index (the winding number) defined as [14]

$$W = \oint_C \frac{d\mathbf{\kappa}}{2\pi} \cdot \left[\frac{h_x}{|\mathbf{h}|}\nabla\left(\frac{h_y}{|\mathbf{h}|}\right) - \frac{h_y}{|\mathbf{h}|}\nabla\left(\frac{h_x}{|\mathbf{h}|}\right)\right].$$

The 2D planar vector **h** around each time reversal related nodal point indeed has vortex structure with nontrivial winding number 1 and -1, while the winding number vanishes once both time reversal partners are included, as is shown in Fig. S2. This is distinct from the Dirac cones splitting out of quadratic degeneracy by symmetry breaking operation in Ref. 25, where the winding number around each is 1 and around the both is 2 instead. This vortex structure in momentum space manifests the topological feature of nodal points.

We now consider symmetry breaking effect on these topological nodal points. An artificial structure without inversion symmetry is shown in Fig. 4(a), where the ratio between $r_1$ and r can be used to describe the degree of inversion symmetry breaking. Here we found that the principal axis of the inversion asymmetric structure orientating parallel to the mirror plane of $m_x$ would lift one pair of topological nodes $N_3$ and $N_4$ while keep the other pair $N_1$ and $N_2$, as is shown in Fig. 4(b) and (c), while a general orientation of the principal axis would break both, as is shown in Fig. 4(d) and (e). This is due to the fact that the former case maintains mirror symmetry of $m_x$, but breaks that of $m_y$, while the latter case both the mirror symmetry of $m_x$ and $m_y$ is broken. Consequently the breaking of mirror symmetry induces hybridization between A and B type bands so that crossing is avoided and topological nodes are unbuckled. This is consistent with the previous analysis that the stability of topological nodal points vanishes as the mirror symmetry protected unavoidable crossing fails. Here, the coexistence of buckled and unbuckled topological nodes is achieved in this photonic system and the width of gap for unbuckled ones is also found to be tunable through varying the degree of inversion asymmetry. This phenomenon gives us the possibility to manipulate the topological nodal points through controlling the mirror symmetry of such inversion asymmetric artificial structure.

*Photonic Chern insulator.* —For photonic crystals with such topological nodes, nodes unbuckling by time reversal symmetry breaking will introduce nontrivial Chern numbers. We use a magnetic permeability tensor with imaginary off-diagonal components to represent the magneto-optical effect and break the time reversal symmetry [26]. It has the following form

$$\hat{\mu} = \begin{pmatrix} \mu & i\mu_\kappa & 0 \\ -i\mu_\kappa & \mu & 0 \\ 0 & 0 & \mu_0 \end{pmatrix}.$$

Introducing magneto-optical effect breaks parity [25], unbuckles all four topological nodes, and brings nonzero Chern number. Since the integration of Berry curvature near each node contributes to a Chern number of ±1/2 [27], the unbuckled four topological nodes in our case give rise to a photonic Chern insulator with Chern number |C|=2. Consistent with the bulk-edge correspondence [28], two different edge states emerge at the boundary of this photonic system, as is shown in Fig. 5 (a).

Photonic Chern insulator with large Chern number becomes realizable. Since the number of equivalent mirror planes in photonic crystal potentially determines how many pairs of topological nodal points can be created, breaking time reversal symmetry by magneto optical effect straightforward combines these with Chern number. It is imaginable that for the triangular lattice in Fig. S1 [17], photonic Chern insulator with three helical edge state is expected.

*Enhancement of finite size effect*. — The helical edge states along the boundary of photonic Chern insulator has unidirectional propagation [2, 8, 29], and the strength of the modes decays exponentially perpendicular to the boundary. Due to the finite size of the sample, the edge mode can leak to the other side and couple with its counterpart to induce a gap in the edge spectrum. This is a typical finite size effect [16]. Interestingly, we find such finite size effect is strengthened by mirror symmetry broken along the propagating direction (here we replace dielectric cylinders with inversion asymmetric artificial structure and align its principal axis along propagating direction). As is shown in Fig. 5(b), there exists a gap in the edge modes spectrum and one edge mode is observed to leak to the other side with longer decay length. Moreover, widening the photonic supercell suppresses the gap in the edge spectrum drastically and reduces the leakage of the corresponding edge mode dramatically, as is shown in Fig. S3. Since the edge spectrum of the two edge modes crosses at different frequencies, such enhanced finite size effect only affect one edge mode at degenerate frequency and does not influence the other. It seems that based on this mechanism, the two helical edge states become controllable and a frequency dependent helical wave filter is feasible.

*Conclusion*. —In summary, we have shown that topological nodal points can potentially split out in photonic crystals with mirror symmetry during the closing of photonic gaps by the touching of two bands with different parity. The full property of topological nodal points, including the emergence and the response to symmetry breaking effect, is then investigated in a photonic square lattice. Such an effect can enable the design of photonic Chern insulator with large Chern number, manipulation of multi-helical edge states, photonic valleytronics, and the development of novel wave functional devices.

## ACKNOWLEDGEMENTS

We thank Professor K. T. Law and Professor Z. Q. Zhang for helpful discussions.

*Corresponding author: phchan@ust.hk

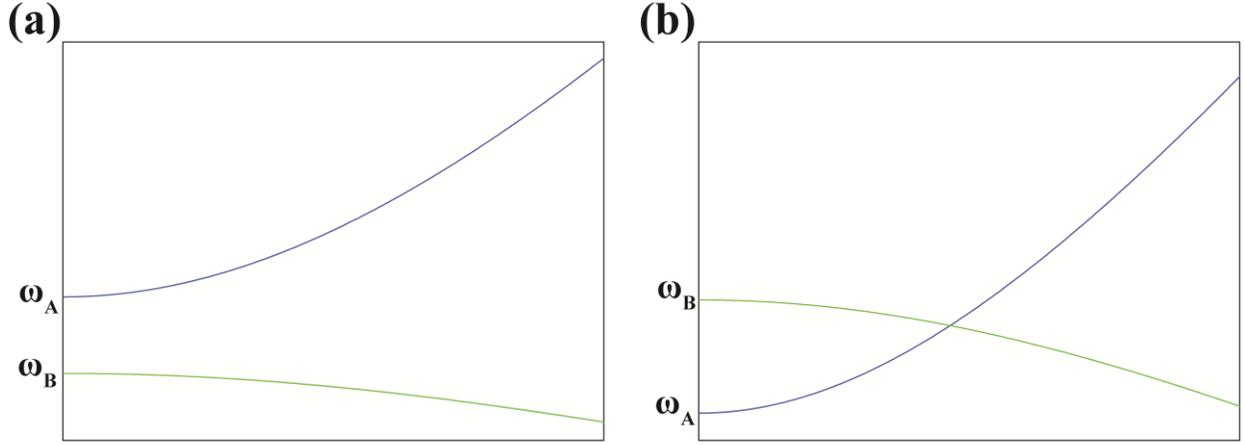

FIG. 1 The schematic diagram of two bands with different parity along a mirror reflection invariant line. In (a) there exists a gap between A type and B type bands, while in (b) the two bands are tuned to have an unavoidable crossing protected by mirror symmetry.

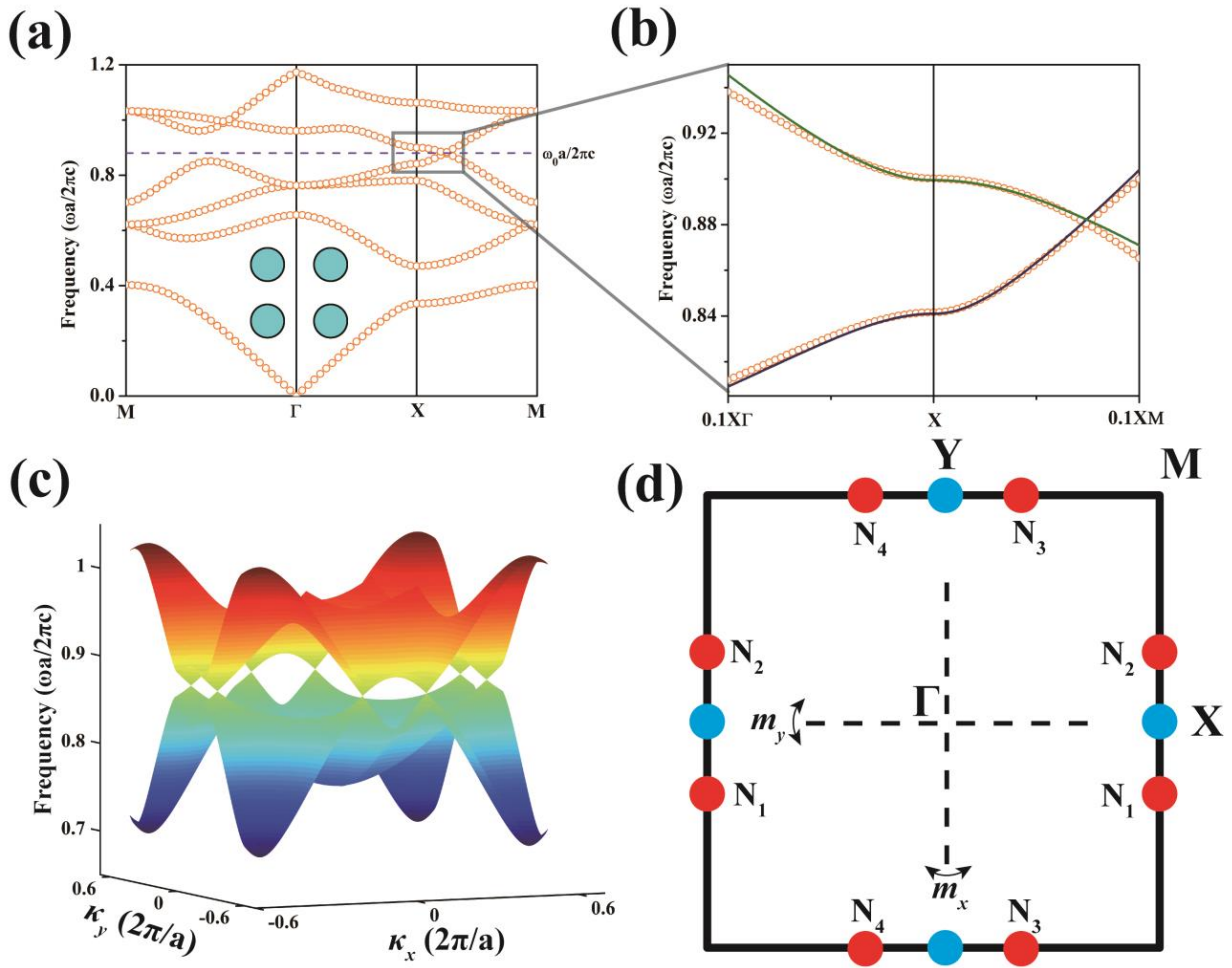

FIG. 2 Band structure of a 2D photonic crystal with a square lattice of dielectric cylinders with relative permittivity $\varepsilon=5.4$ and radius $r=0.2a$ embedded in air. Here $a$ is the lattice constant. (a) The band structure along high symmetry lines. Gapless topological nodal points with linear dispersions at frequency $\omega_0$ emerge along XM. (b) Enlarged view near the nodal point, with the solid lines showing k·p perturbation result. (c) Three dimensional dispersion surface containing four topological nodal points in the first Brillouin zone. (d) The red dots mark the positions of topological nodal points $\{N_1, N_2, N_3, N_4\}$ in the first Brillouin zone. The two perpendicular dot lines represent the mirror operation $m_x$ and $m_y$.

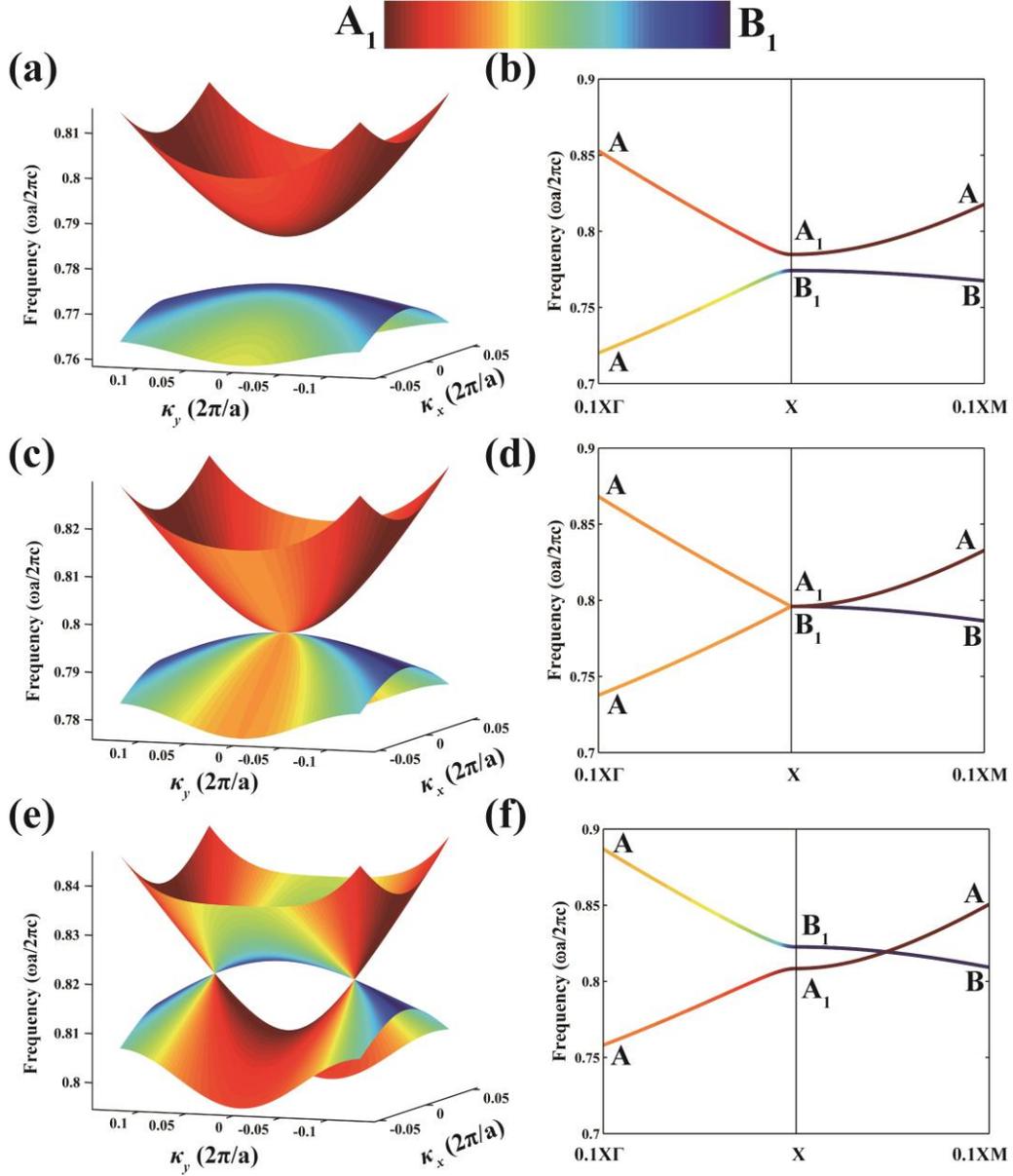

FIG. 3 Evolution of A1 and B1 subbands with variation of cylinders' relative dielectric constant showing the emergence of topological nodal points as the photonic band gap closes upon a reduction of the dielectric constant. The relative permittivity and radii adopted are {ε=8.8, r=0.2a} for (a) and (b), {ε=8, r=0.2a} for (c) and (d), {ε=7.2, r=0.2a} for (e) and (f). The colored shading indicates the symmetry type of the band at the corresponding point in Brillouin zone. A and B are irreducible representation of $C_{1h}$ group with even and odd parity respectively.

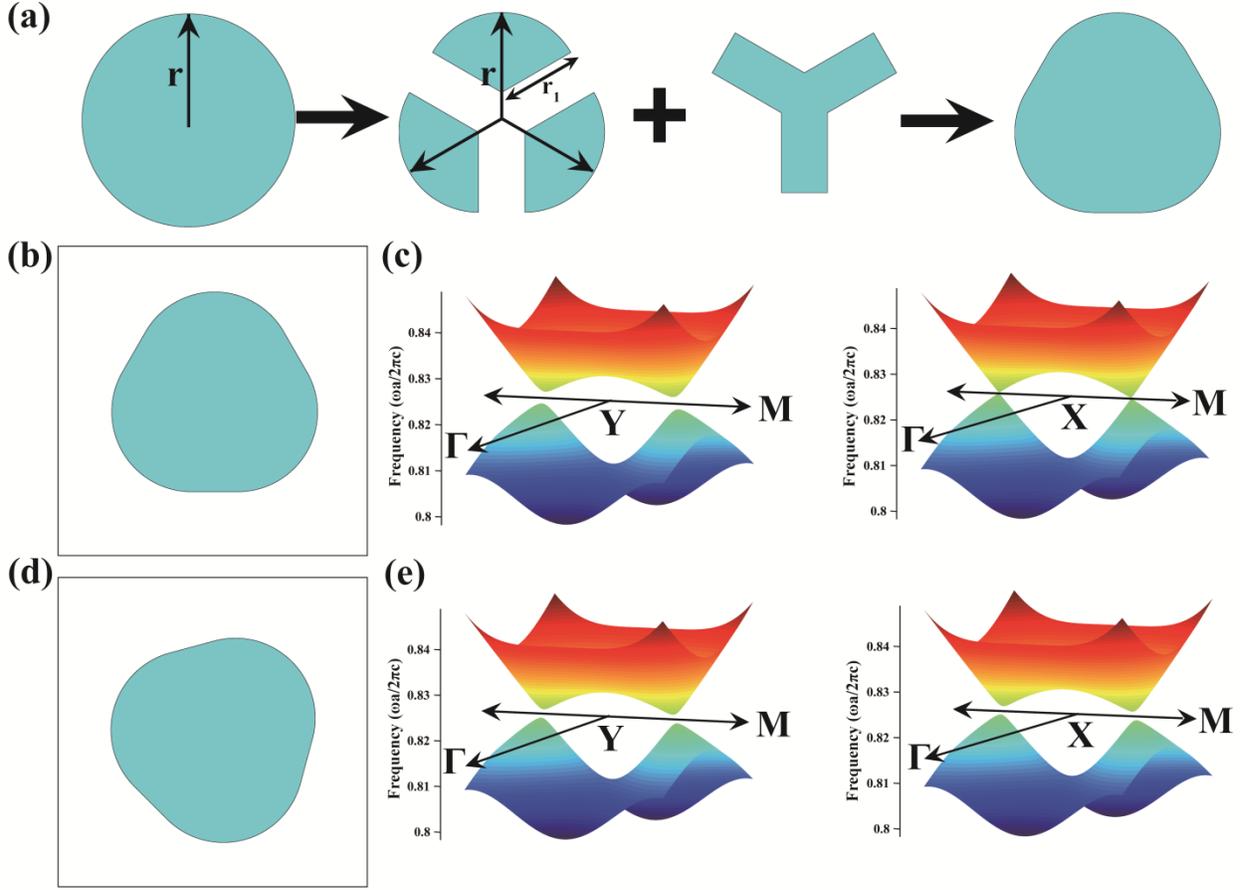

FIG. 4 The effect of inversion symmetry breaking on the band structure. (a) Design of an artificial structure with a smooth transition from inversion symmetry to lack of inversion. First, the cylinder is narrowed and trisected. Then the empty area between the three sectors is filled with the same material of geometry endowing a smooth profile of the whole structure. For (b) the inversion asymmetrical artificial structure of {$\varepsilon=7.2$, $r_1=0.18a$, $r=0.2a$} in the unit cell aligns its principal axis parallel to a lattice unit vector while it is rotated by 45° for (d). (c) The band structure near lifted degenerate points around Y and maintaining topological nodal points around X. (e) The band structure near lifted degenerate points around both Y and X.

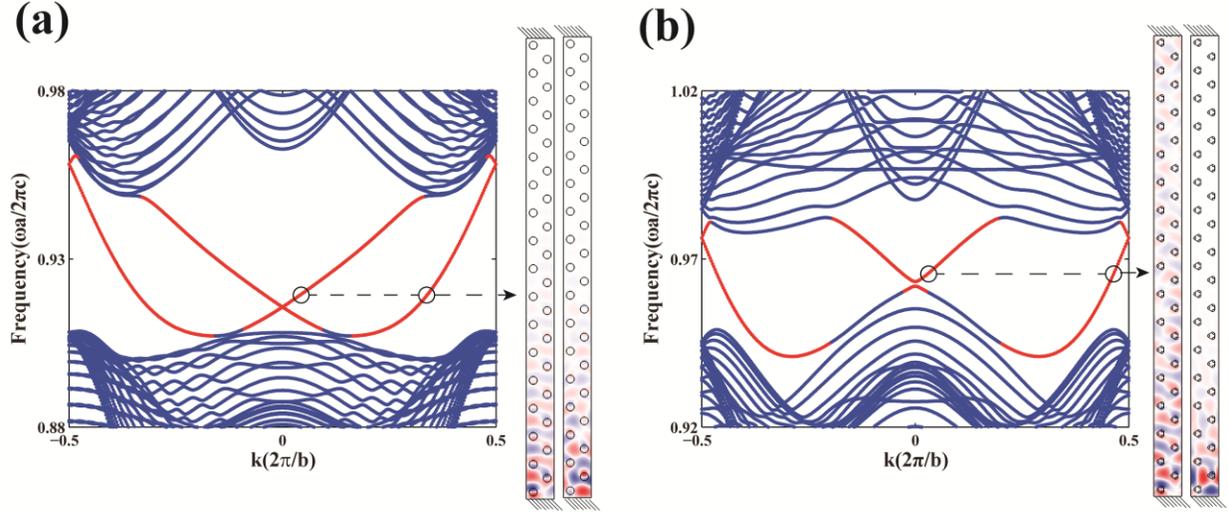

FIG. 5 Topological edge states induced by magneto-optical effect. The projected frequency spectrum of the photonic crystal ribbon containing 33 layers with PEC boundary (the shaded area at both ends means the PEC boundary) in the end is shown with structure parameters {$r_1$=0.2a, r=0.2a} for (a) and {$r_1$=0.1a, r=0.2a} for (b). The inset on the right shows the electric field distribution of the two edge modes marked by circle arranged in the same order. In (b), the inversion asymmetry along the helical propagating direction strengthens the finite size effect and opens a gap in edge mode spectrum. In both cases ε=5.4, μ=$μ_0$=1, and $μ_κ$=0.5 are taken. Here the periodicity of the ribbon is denoted by $b=\sqrt{2}a$.

**Supplementary Information for**
**The Emergence of Topological Nodal Points in Photonic Crystal with Mirror Symmetry**
Wen-Yu He and C. T. Chan[*]

Department of Physics and Institute for Advanced Study, Hong Kong University of Science and Technology, Clear Water Bay, Kowloon, Hong Kong, China

I. Derivation of the effective Hamiltonian using symmetry constraints

In general, the 2×2 effective Hamiltonian is given by

$$H_{\text{eff}}(k_{\|},k_{\perp}) = a(k_{\|},k_{\perp})\sigma_0 + b(k_{\|},k_{\perp})\sigma_x + c(k_{\|},k_{\perp})\sigma_y + d(k_{\|},k_{\perp})\sigma_z$$

Under the mirror operator, the effective Hamiltonian must behave like the following

$$M H_{\text{eff}}(k_{\|},k_{\perp}) M^{-1} = H_{\text{eff}}(k_{\|},-k_{\perp})$$

On the basis of eigenfunction with different parity, the matrix representation of mirror operator is $\sigma_z$. Thus along the mirror reflection invariant line, $\sigma_z H_{\text{eff}}(k_{\|}) \sigma_z^{-1} = H_{\text{eff}}(k_{\|})$ gives

$$a(k_{\|})\sigma_0 + b(k_{\|})\sigma_x + c(k_{\|})\sigma_y + d(k_{\|})\sigma_z = a(k_{\|})\sigma_0 - b(k_{\|})\sigma_x - c(k_{\|})\sigma_y + d(k_{\|})\sigma_z$$

It makes $\sigma_x$ and $\sigma_y$, the off diagonal terms vanish. Considering the perpendicular component, it gives that

$$a(k_{\|},k_{\perp})\sigma_0 + b(k_{\|},k_{\perp})\sigma_x + c(k_{\|},k_{\perp})\sigma_y + d(k_{\|},k_{\perp})\sigma_z = a(k_{\|},-k_{\perp})\sigma_0 - b(k_{\|},-k_{\perp})\sigma_x - c(k_{\|},-k_{\perp})\sigma_y + d(k_{\|},-k_{\perp})\sigma_z$$

Expanding them near the crossing point to the first order and following the restriction that $b(k_{\|}, k_{\perp})$ and $c(k_{\|}, k_{\perp})$ are both odd in $k_{\perp}$, we obtain

$$a(k_{\|},k_{\perp}) = \frac{1}{2}(v_A + v_B)k_{\|}$$
$$b(k_{\|},k_{\perp}) = (v_+ + v_-)k_{\perp}$$
$$c(k_{\|},k_{\perp}) = (v_+ - v_-)k_{\perp}$$
$$d(k_{\|},k_{\perp}) = \frac{1}{2}(v_A - v_B)k_{\|}$$

The eigen value of this effective Hamiltonian reads

$$\Delta\omega = \frac{1}{2}\left[(v_A + v_B)k_{\|} \pm \sqrt{(v_A - v_B)^2 k_{\|}^2 + 8(v_+^2 + v_-^2)k_{\perp}^2}\right]$$

II. Symmetry analysis with group theory in the **k·p** perturbation

The effective Hamiltonian matrix is mainly decided by the coefficient ($p_{ljx}$, $p_{ljy}$) and $q_{lj}$ [S1]. The magnetic permeability of the cylinders is assumed to be one and then the expression of $\mathbf{p}_{lj}$ and $q_{lj}$ can be simplified as

$$p_{ljx} = i\frac{(2\pi)^2}{\Omega} \int \psi^*_{l\mathbf{k}_0}(\mathbf{r}) \cdot 2\frac{\partial}{\partial x} \psi_{j\mathbf{k}_0}(\mathbf{r}) d\mathbf{r}$$

$$p_{ljy} = i\frac{(2\pi)^2}{\Omega} \int \psi^*_{l\mathbf{k}_0}(\mathbf{r}) \cdot 2\frac{\partial}{\partial y} \psi_{j\mathbf{k}_0}(\mathbf{r}) d\mathbf{r}$$

$$q_{lj} = i \frac{(2\pi)^2}{\Omega} \int \psi_{l\mathbf{k}_0}^*(\mathbf{r}) \psi_{j\mathbf{k}_0}(\mathbf{r}) d\mathbf{r}$$

The operator $\frac{\partial}{\partial x}$ behaves like $B_1$ while $\frac{\partial}{\partial y}$ behaves like $B_2$ in $C_{2v}$ group. Combining this fact with the direct product table for $C_{2v}$ group, we can find the matrix elements obtained in the effective Hamiltonian are the only way to make all become $A_1$, the full symmetric representation. This indicates that such style of effective Hamiltonian is symmetry decided.

## III. Band dispersion along XM and ΓX

Along XM and ΓX, the effective Hamiltonian matrix is block diagonalized. The eigen frequency dispersion can be solved analytically. For $\kappa_x=0$ and along $\kappa_y$, we have

$$\frac{\tilde{\omega}_{B_1}^2}{c^2} = \frac{1}{2}\left[\left(\frac{\omega_{B_1}^2+\omega_{A_2}^2}{c^2}+q_{B_1B_1}\kappa_y^2+q_{A_2A_2}\kappa_y^2\right) - \sqrt{\left(\frac{\omega_{A_2}^2-\omega_{B_1}^2}{c^2}+q_{B_1B_1}\kappa_y^2-q_{A_2A_2}\kappa_y^2\right)^2+4p_{A_2B_1}^2\kappa_y^2}\right]$$

and

$$\frac{\tilde{\omega}_{A_1}^2}{c^2} = \frac{1}{2}\left[\left(\frac{\omega_{B_2}^2+\omega_{A_1}^2}{c^2}+q_{B_2B_2}\kappa_y^2+q_{A_1A_1}\kappa_y^2\right) + \sqrt{\left(\frac{\omega_{A_1}^2-\omega_{B_2}^2}{c^2}+q_{B_2B_2}\kappa_y^2-q_{A_1A_1}\kappa_y^2\right)^2+4p_{A_1B_2}^2\kappa_y^2}\right].$$

Since the band order of $\{B_1, A_2\}$ and $\{A_1, B_2\}$ is fixed, these two bands would inevitably cross once $\omega_{A_1}$ becomes lower than $\omega_{B_1}$. For $\kappa_y=0$ and along $\kappa_x$, we have

$$\frac{\omega_\pm^2}{c^2} = \frac{1}{2}\left[\left(\frac{\omega_{A_1}^2+\omega_{B_1}^2}{c^2}+q_{B_1B_1}\kappa_x^2+q_{A_1A_1}\kappa_x^2\right) \pm \sqrt{\left(\frac{\omega_{B_1}^2-\omega_{A_1}^2}{c^2}+q_{B_1B_1}\kappa_x^2-q_{A_1A_1}\kappa_x^2\right)^2+4p_{A_1B_1}^2\kappa_x^2}\right].$$

It is easy to see that $\tilde{\omega}_{A_1} = \omega_+$ if $\omega_{A_1} > \omega_{B_1}$ and $\tilde{\omega}_{B_1} = \omega_+$ if $\omega_{B_1} > \omega_{A_1}$, meaning that band inversion occurs in this direction once the two bands touch.

## IV. Effective two-band model

Since $A_2$ and $B_2$ bands are far from the degeneracy frequency, we take a procedure to reduce the effective Hamiltonian to a two-band model. Here, we drop $A_2$ and $B_2$ bands off from the effective Hamiltonian, but correct the two-band model through virtual process in which electromagnetic mode jumps from $A_1$ or $B_1$ band to $A_2$ or $B_2$ band, then back to the original band [S2, S3]. Thus the eigen problem can be written as

$$H_{11}\psi_{B_2} + H_{12}\psi_{A_1} + H_{14}\psi_{A_2} = \frac{\omega_0^2}{c^2}\psi_{B_2} \quad (1)$$

$$H_{21}\psi_{B_2} + H_{22}\psi_{A_1} + H_{23}\psi_{B_1} = \frac{\omega^2}{c^2}\psi_{A_1} \quad (2)$$

$$H_{32}\psi_{A_1} + H_{33}\psi_{B_1} + H_{34}\psi_{A_2} = \frac{\omega^2}{c^2}\psi_{B_1} \quad (3)$$

$$H_{41}\psi_{B_2} + H_{43}\psi_{B_1} + H_{44}\psi_{A_2} = \frac{\omega_0^2}{c^2}\psi_{A_2} \quad (4)$$

Here $H_{ij}$ is the element of original four by four effective Hamiltonian matrix, and $\omega_0$ is the degenerated frequency, which can be obtained through the equation $\tilde{\omega}_{A_1} = \tilde{\omega}_{B_1}$. Thus by expressing $\psi_{A_2}$, $\psi_{B_2}$ in terms of $\psi_{A_1}$, $\psi_{B_1}$, and substituting them back into Eq. (2) and (3), the two-band Hamiltonian matrix is determined as

$$\tilde{H} = \begin{pmatrix} H_{22} & H_{23} \\ H_{32} & H_{33} \end{pmatrix} + \frac{1}{\left(H_{11} - \frac{\omega_0^2}{c^2}\right)\left(H_{44} - \frac{\omega_0^2}{c^2}\right) - H_{41}H_{14}} \begin{pmatrix} -H_{21}\left(H_{44} - \frac{\omega_0^2}{c^2}\right)H_{21} & H_{21}H_{14}H_{43} \\ H_{12}H_{41}H_{34} & -H_{34}\left(H_{11} - \frac{\omega_0^2}{c^2}\right)H_{43} \end{pmatrix}$$

Checking the matrix element we can confirm that the diagonal terms are real and off-diagonal terms are imaginary, and all terms are functions of ($\kappa_x$, $\kappa_y$). Therefore, setting

$$\tilde{H} = \begin{pmatrix} a(\kappa_x, \kappa_y) & -ib(\kappa_x, \kappa_y) \\ ib(\kappa_x, \kappa_y) & c(\kappa_x, \kappa_y) \end{pmatrix}$$

we can expand the two-band model in the basis of Pauli matrx as follows

$$\tilde{H} = \frac{1}{2}\left(a(\kappa_x, \kappa_y) + c(\kappa_x, \kappa_y)\right)I + b(\kappa_x, \kappa_y)\sigma_y + \frac{1}{2}\left[a(\kappa_x, \kappa_y) - c(\kappa_x, \kappa_y)\right]\sigma_z.$$

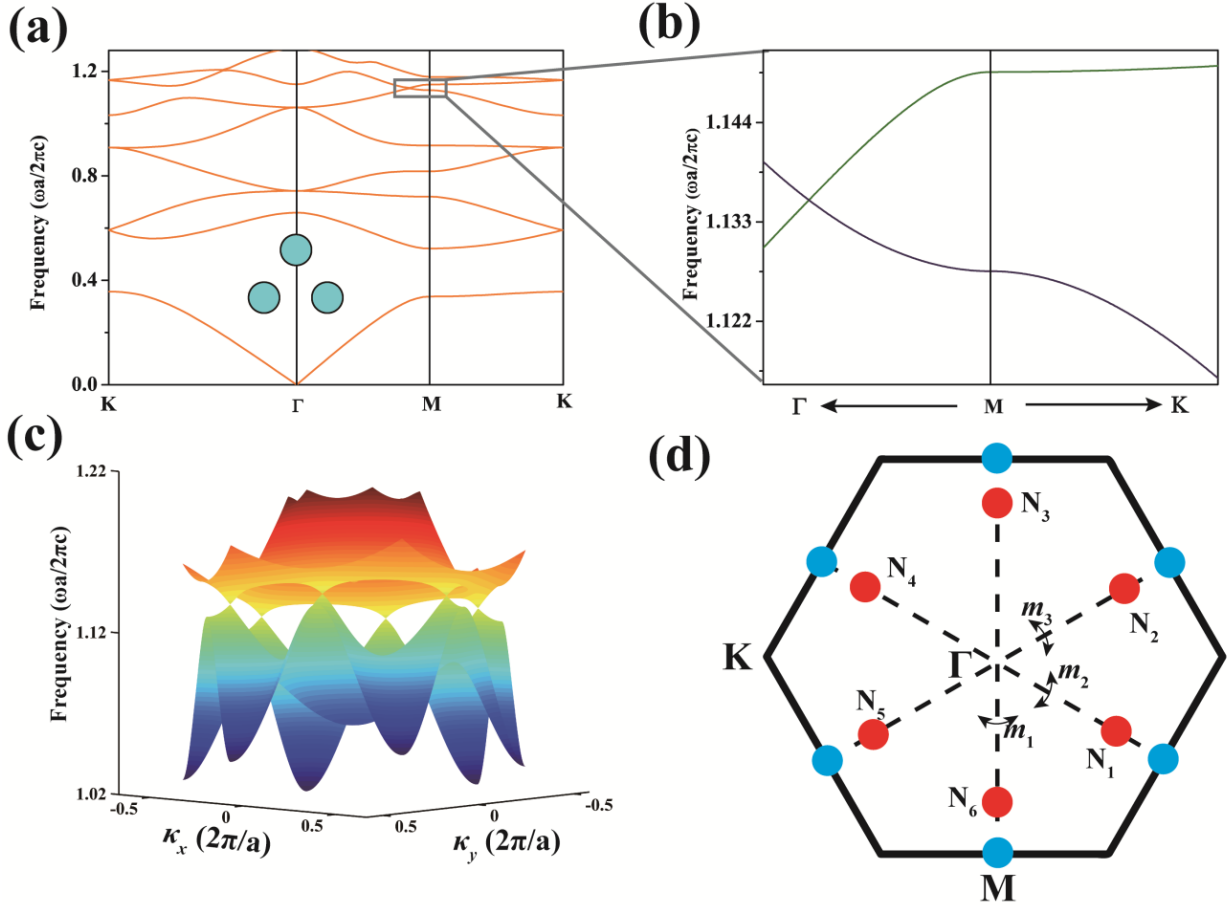

FIG. S1 Band structure of a 2D photonic crystal with a triangular lattice of dielectric cylinders with relative permittivity ε=7.6 and radius r=0.18a embedded in air. Here a is the lattice constant. (a) The band structure along high symmetry lines. Gapless topological nodal points with linear dispersions emerge along ΓM. (b) Enlarged view near the nodal point. (c) Three dimensional dispersion surface containing four topological nodal points in the first Brillouin zone. (d) The red dots mark the positions of topological nodal points {$N_1$, $N_2$, $N_3$, $N_4$, $N_5$, $N_6$} in the first Brillouin zone. The three dot lines represent the mirror operation $m_1$, $m_2$ and $m_3$.

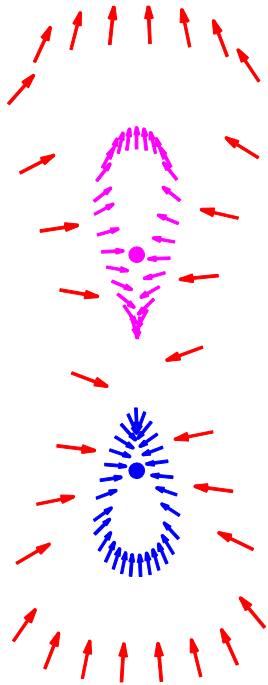

FIG. S2 (a) Topological nodal points characterized by topological vortex. The winding number of time reversal related topological vortices is opposite to each other as shown.

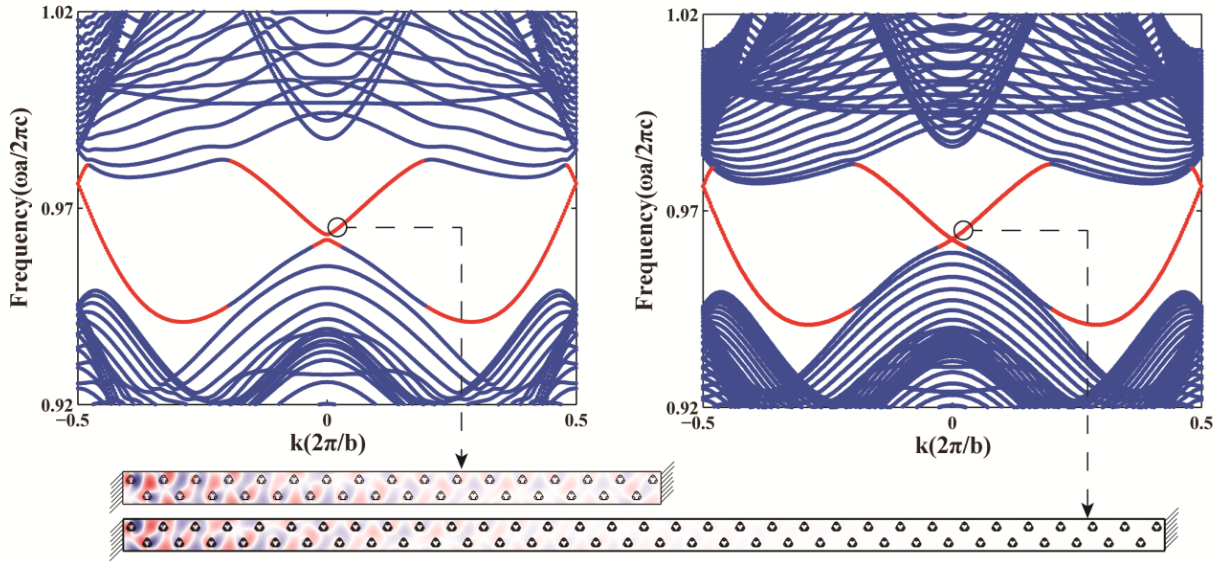

FIG. S3 The finite size effect upon the edge modes. The gap induced by finite size effect in edge modes spectrum is suppressed dramatically by widening the photonic crystal ribbon, consistent with the difference between edge spectrum in right panel and left panel. The corresponding electric field distribution of these edge modes are shown at the bottom.